# Critical Evaluation of Relative Importance of Stress and Stress Gradient in Whisker Growth in Sn Coatings

Piyush Jagtap, Vijay A. Sethuraman, and Praveen Kumar[*]
Department of Materials Engineering, Indian Institute of Science, Bangalore-560012, India.

**Abstract**
Here, we study the role of stress state and stress gradient in whisker growth in Sn coatings electrodeposited on brass. The bulk stress in Sn coatings was measured using a laser-optics based curvature setup, whereas glancing angle X-ray diffraction was employed to quantify the surface stress; this also allowed studying role of out-of-plane stress gradient in whisker growth. Both bulk stress and surface stress in Sn coating evolved with time, wherein both were compressive immediately after the deposition, and thereafter while the bulk stress monotonically became more compressive and subsequently saturated with aging at room temperature, the stress near the surface of the Sn coating continually became more tensile with aging. These opposing evolutionary behaviors of bulk and surface stresses readily established a negative out-of-plane stress gradient, required for spontaneous growth of whiskers. The importance of out-of-plane stress gradient was also validated by externally imposing widely different stress states and stress gradients in Sn coatings using a 3-point bending apparatus. It was consistently observed that whisker growth was more in the coatings under external tensile stress, however, with higher negative out-of-plane stress gradient. The results conclusively indicate the critical role of negative out-of-plane stress gradient on whisker growth, as compared to only the nature (i.e., sign and magnitude) of stress.

## 1.  Introduction

Whisker growth in electrodeposited Sn coatings is a major concern in Pb-free Sn platings and solders used in various electronic applications. Sn whiskers, which are filament type Sn structures with very high aspect ratio (e.g., $1:10^4$ [1]), readily cause short-circuiting when it connects two adjacent current carrying conductors. It is therefore of great scientific and technological interest to understand their growth mechanism and develop effective mitigation strategies to prevent their growth during the service life of electronic devices. It is now understood that whisker growth in Sn coating is a diffusion-controlled stress relaxation phenomenon [2], wherein Sn atoms are incessantly transported from the bulk of the coating to the root of whiskers [2-7]. Moreover, it is also generally considered that compressive stress in the Sn coatings, which is continuously regenerated due to the growth of interfacial intermetallic compound (IMC) $Cu_6Sn_5$, is responsible for the growth of whiskers [6, 7]. In fact, mass transport of atoms from the highly compressed region to the stress-free whisker root during whisker growth and an accelerated whisker growth rate under externally applied compressive stress were reported very early in 1954 [2]. The stresses, both internal [3, 4] and external [2, 5], and the stress gradients, both in-plane and out-of-plane [6], have been suggested to play crucial role in whisker growth. Nevertheless, as explained below, a detailed study delineating the importance of the stress gradient and the nature (i.e., sign and magnitude) of the stress on Sn whiskering is still not available in the open literature.

---

[*]Corresponding author (PK): E-mail: praveenk@iisc.ac.in, Tel.: +91-80-2293 3369

The first set of evidences revealing the role of stress in Sn whiskering came from classic X-ray diffraction (XRD) based studies [7-13], which showed changes in the lattice spacing of the Sn coating and the Cu substrate such that Sn and Cu layers were under compressive and tensile stresses, respectively [7]. The compressive stress in Sn coatings was attributed to the inter-diffusion and reaction between Cu and Sn that produced interfacial $Cu_6Sn_5$ IMC layer in between the Sn coating and the substrate. In this context, a few interesting observations were made: (i) placing a thin Ni layer between Sn and Cu substrate, which completely stopped whisker growth, prevented Sn coating from becoming compressive [8], (ii) existence of a threshold compressive stress (e.g., 45 MPa, externally applied [4]) below which whiskers did not grow, (iii) "baking" Sn coatings at 150 ºC for 1 h, which lead to prevalence of tensile stress in Sn coatings, prevents growth of whiskers even after long duration (e.g., 700 h [12]), and (iv) the electrolytic baths that produces Sn coatings having more tensile stress (e.g., non-methanesulphonic acid (MSA) as compared to MSA bath) were effective in reducing overall whisker growth [9]. Therefore, it became a thumb rule to solely correlate whiskering with compressive stress in Sn coating. Nevertheless, contrary reports are also available in literature. For example, Lal and Moyer [11] used XRD based conventional $sin^2\psi$ method to quantify the stresses in Sn coatings deposited using various bath chemistries and current densities. However, they did not observe a consistent relationship between the residual stress and the whisker growth. This may be partly because various Sn coatings were of different thickness and they also had different crystallographic textures, which are known to significantly affect the whiskering [10, 11]. However, as it will be evident from the results obtained in this study, the above may also be due to the lack of information on out-of-plane stress gradient.

The other set of studies illustrating the role of stress came from determination of stresses by measuring the curvature of coating-substrate system [10, 14-20]. Although this method does not elucidate the mechanism of stress evolution, its simplicity proffers simultaneous measurement of residual stress and the whisker density in real time. Thus, a direct correlation between the stress state and the whisker growth rate could be established, wherein it has been shown that stress in Sn coatings rapidly approached the maximum compressive stress and due to the whisker growth, the stress became tensile or saturated at a smaller compressive value [10, 14, 16-20]. Curvature measurement techniques also showed the efficacy of post deposition "baking" of the coating at 150 ºC for 1 h [9] as well as alloying Sn with Pb [14, 17, 18] in preventing the build-up of compressive stress in the Sn coatings, eventually resulting in whisker-free Sn coatings.

In addition, contradictory to many reports, whiskers were also observed to grow in Sn coatings under tensile stress, such as in reflowed Sn coating [21-22], and in Sn-Mn alloy coating [15]. It was observed that whisker growth in Sn-Mn alloy was observed within a few hours after the deposition and coating remained under tensile stress during entire observation period of whisker growth [15]. Furthermore, it was reported that the Sn coatings baked at 150 °C for 1 h developed compressive stress after a prolonged aging (e.g., 150 days or 3600 h); however, these coatings still remained whisker-free [13]. These "anomalous" observations suggest that the nature of stress alone may not be the key factor driving whisker growth.

In addition to studying effect of stress on whisker growth, a few research groups also investigated the role of stress gradient; however, the literature exploring the effect of stress gradient on whisker growth is rather scantier [6, 13, 23-24]. The only known experimental study [13] employing XRD-based technique for measuring stress as function of distance from the surface reported that negative out-of-plane stress gradient is necessary for whisker growth. However, since X-rays have limited depth of penetration in Sn coatings, the stress gradient through this method could be measured only in the top 1 μm of a 4 μm thick coating. It should be noted that growth of IMC layer, which is often attributed to the generation of compressive stress in Sn coatings showing whisker growth, occurs at the interface of Sn and substrate, which is often far away from the reach of X-rays. This has been a common shortcoming of all the previous studies [8, 9, 12] using XRD techniques for measurement of stress in Sn coating, as they are not true representative of entire Sn coating, especially as the stress gradient exists in Sn coatings across the entire thickness. In a later report using synchrotron micro-diffraction for studying local stress field surrounding whisker [6], it was reported that both in-plane and out-of-plane stress gradients are essential for whisker growth.

In the context of evaluating the importance of stress, the role of external stress has also been studied in the past by several research groups [2, 5, 25-30]; however, contradictory results were often reported. For example, a few studies reported accelerated growth of whiskers under applied compressive stress [2,5,25], while a few other reported the contrary [26]. Furthermore, a few studies reported that the whiskers grown under compression were longer and more in number than the stress-free region of the sample [27, 30], while a few reported that whisker growth was the highest in the Sn coating under tensile stress, as well as the length of the whiskers in the coatings under tensile stress was longer compared with length of whiskers in the compressively stressed Sn coatings [28, 29]. However, explicit explanation of the above "apparently anomalous" results or dichotomy in the results were not provided in any of the studies.

It is evident from the aforementioned discussion on stress measurement in Sn coatings and the effect of stress and stress gradient on whisker growth that the role of stress in whisker growth is controversial. Especially, the growth of whiskers under tensile stress remains anomalous and has never been systematically addressed in previous works. In addition, above discussion also indicates that the experimental evidence confirming (or refuting) the importance of stress gradient and the nature of stress (i.e., compressive or tensile as well as magnitude) on the whisker formation remains scantier. Accordingly, this study aims to critically examine the relative importance of the nature of stress and the stress gradient on Sn whiskering by quantifying the evolution in the residual stresses across the thickness of Sn coatings during aging and by imposing widely different external stresses and stress gradients on Sn coatings, and thus aims to systematically develop a fundamental understanding about whisker growth.

## 2. Experimental Materials and Procedure

### 2.1 Stress measurement using substrate curvature measurement

Brass coupons of size $2 \times 2$ cm$^2$ and thickness of 300 μm were cut out from a large sheet using wire electro discharge machining (W-EDM). Both surfaces of the brass coupons were then metallographically polished up to 0.05 μm surface finish. Subsequently, a coupon of $1 \times 1.5$ cm$^2$ size was sectioned from the central region of the polished coupon using W-EDM at slow speed.

This produced a reasonably flat and highly reflective surface required for measuring stress in the Sn coating using an optical technique for measuring wafer curvature.

The curvature of the substrate was recorded using a setup based on laser optics. For details of the experimental setup, one can refer to [31-33] as here only important aspects are described for continuity. The wafer curvature measurement setup was placed on a vibration isolation optical table to preclude any error due to vibrations over very long periods of observation. The setup comprised a He-Ne gas laser source, a beam splitter, an optically flat mirror and a charge-coupled device (CCD) detector connected to a computer that allowed continuous monitoring of the substrate curvature. The inclination of the beam splitter, which was placed in the path of the incident laser beam, relative to the laser beam was controlled to obtain an array of multiple laser spots, and the spacing between the laser spots after being reflected by the substrate or wafer was continuously monitored in real time using the CCD detector. The curvature, κ, was calculated by measuring the fractional change in the laser spot spacing, and subsequently the film stress, σ'$_f$, was estimated using Stoney's equation as follows:

$$\sigma'_f = \frac{E_s}{6(1-\nu_s)} \frac{d_s^2 \kappa}{d_f} \qquad (1)$$

where $E_s$ and $\nu_s$ are the Young's modulus and Poisson's ratio of substrate, respectively, and $d_s$ and $d_f$ are thicknesses of the substrate and the film, respectively. However, it should be noted that Stoney's formula, as given in **Eq. 1**, is strictly valid only if the substrate is absolutely flat prior to the film deposition, which is often not the case. Therefore, change in the curvature (Δκ) was calculated from curvature of the substrate before and after the electro-deposition of Sn coating. This necessitated that the value of the curvature, κ, in Stoney's equation, as given in **Eq. 1**, should be replaced by change in curvature, Δκ. This allowed measurement of a change in the stress in the film. It should be noted that the IMC layer formed in between Sn and brass was discontinuous and hence, as it will be discussed later, the change in curvature due to discontinuous IMC layer would be negligible [34, 35]. Therefore, effect of the IMC layer on the stress in Sn coating was neglected.

After measuring the initial curvature of the polished brass coupons, 4 μm thick Sn was coated on it using electro-deposition using acidic stannous sulfate bath. Prior to electrodeposition brass coupons were cleaned using isopropyl alcohol followed by acid rinsing to remove surface oxide layer. Pure Sn (99.99 % purity) was used as an anode while brass coupons were attached to cathode of an electrolytic cell. The electrolytic solution was stirred using a magnetic needle at 200 rotations per minute to maintain uniform hydrodynamic conditions. The electro-deposition was performed at constant bath temperature of 40 °C using a current density of 20 mA/cm$^2$ for 330 s. Sn was selectively deposited on only one side of the brass coupons by masking the other side using Kapton® tape. Following electro-deposition, the Kapton® tape was removed by dipping the samples in acetone. The change in the curvature of the substrate was then continuously monitored in real time using the laser-optics based system. It should be noted that the laser reflected from the backside of the brass coupon, where Sn was not coated.

## 2.2 Stress measurement using X-ray diffraction

The stress in the vicinity of the surface of the Sn coating was measured using another set of representative samples, deposited at the same bath temperature and the current density as mentioned in **Section 2.1**, using an X-ray diffractometer (Bruker TXS D8 Discover) with a parallel beam geometry. The turbo X-ray source (TXS) rotating anode produced highest intensities of X-rays that allowed measuring shift in high angle peaks that are not captured in ordinary X-ray diffractometers operating at lower X-ray currents. Cu K$_\alpha$ radiation from a rotating anode operating at 45 kV and 100 mA was used in this study. 0.2 mm incident slit was used to restrict the beam size of X-rays. The sample was mounted on Eulerian cradle, which allowed the sample tilts and rotation. The stress measurement was performed in ω-mode (or glancing angle mode) [36], which allowed measurement of the inter-planar spacing of set of (*h k l*) planes that are inclined to sample surface. The angle of incidence, α, was varied from 1-10°, thus changing the effective penetration depth of the X-rays. The penetration depth of the X-ray beam, τ, can be calculated from the incident (α) and the exit angles (β), as follows [37]:

$$\tau = \frac{\sin\alpha \sin\beta}{\mu(\sin\alpha + \sin\beta)} \quad (2)$$

where μ is the mass absorption coefficient (μ/ρ equal to 274 cm$^{-1}$ for Sn [38], where ρ is the density of Sn). Thus, the variation in incidence angle proffered probing strain in (501)-planes at different penetration depths, ranging from 0.1 to 0.76 μm in a 4 μm thick coating, from the free surface of the Sn coatings. Therefore, the stress was measured up to a penetration depth of only top ~15 % of the total thickness of the coating and hence reasonably representing average stress in the top layer of the coating or surface stress.

As mentioned above, the diffraction peak corresponding to (501)-reflection of Sn was collected at various α and a shift in the position of peak was monitored. The peak measurement was performed with step size of 0.0005 and with a step time of 3 s. Therefore, measurement of one peak took ≈2 h and the entire measurement took ≈20 h. The peak positions of (501)-reflection at different α were calculated by peak fitting procedure in Origin® software using Pearson VII peak function [37]. The average stress was calculated from linear fit between $\varepsilon^{\psi\varphi}_{hkl}$ (which is $\psi\varphi$-component of the strain corresponding to (*hkl*)-plane) and $sin^2\psi$, as dictated by the following relationship which relates strain measured using XRD to the mechanical stress tensor for macroscopically isotropic specimen under rotationally symmetric plane stress (i.e., $\sigma_{11} = \sigma_{22} = \sigma_f$ and $\sigma_{12}, \sigma_{13}, \sigma_{23}$ and $\sigma_{33}=0$) condition:

$$\varepsilon^{\psi\varphi}_{hkl} = \left(\frac{1}{2} S_2^{hkl} \sin^2\psi + 2S_1^{hkl}\right)\sigma_f \quad (3)$$

where $S_1^{hkl}$ and $S_2^{hkl}$ are diffraction elastic stiffness constants. Strain-free lattice spacing of (501)-reflection given in Joint Committee on Powder Diffraction Standards (JCPDS) was used to calculate the strain using the following equation.

$$\varepsilon^{\psi\varphi}{}_{hkl} = \frac{d_{hkl} - d_0}{d_0} \tag{4}$$

## 2.3 Application of external stress and stress gradient using three-point bending

To study the effect of external stress and stress gradient on whisker growth, a 3-point bending setup, shown in **Figure 1,** was used. Firstly, Sn coatings were electro-deposited, using the process described earlier in **Section 2.1** with the same bath temperature and current density, on only one side of a 200 μm thick and 5 cm long brass strip. **Fig. 1** schematically illustrates the strategies employed for application of external compressive and tensile stresses onto the samples. The advantages of such a setup are: (i) it allowed quantifying the imposed stress and stress gradients using simple finite element analysis (FEA) of a bimetallic strip under bending load, (ii) tensile and compressive stress of the same magnitude can be imposed in the coatings on the opposite sides of the sample at the same time, as shown schematically in **Fig. 1**, (iii) data corresponding to various stresses can be obtained using one experiment, and (iv) bending of Sn coating in such a manner also produced stress gradients along the thickness of the coating. In particular, as shown in **Fig. 1b**, the application of stress in such a way created the stress gradients along the thickness direction, such that the surface of the Sn coating was at maximum tension or compression, depending on the configuration chosen in **Fig. 1a**, as compared to the Sn atoms at the Sn coating-brass interface.

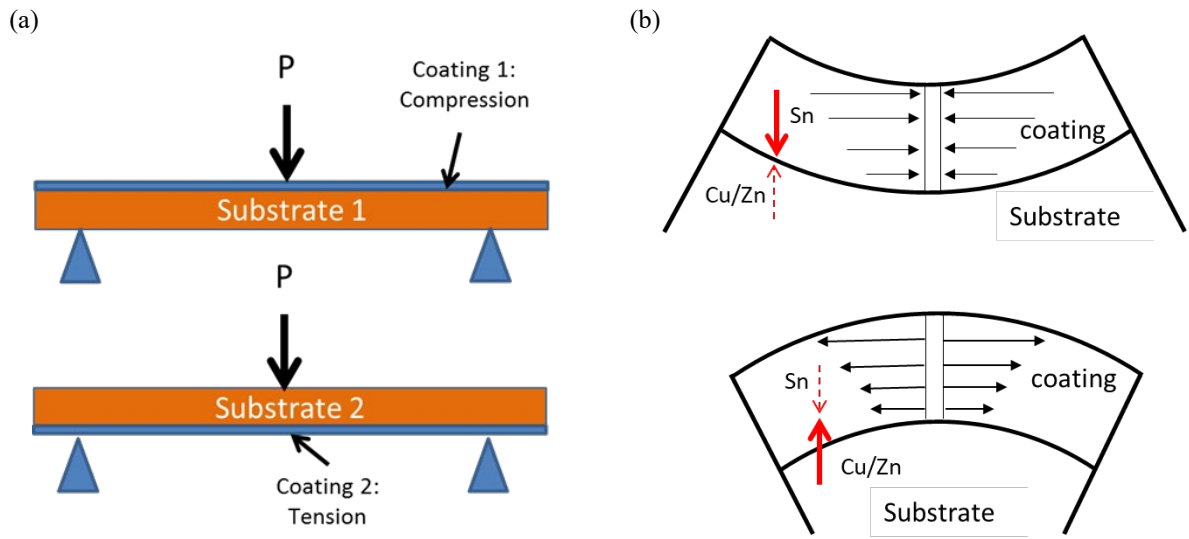

Fig. 1: (a) 3-point bending scheme used to apply external stress and stress gradient on the Sn coated brass. (b) Schematic illustration of the stress gradient produced as per the schematic shown in (a). The size and the direction of horizontal arrows in (b) denote the magnitude and sign of the dominant stress component in the Sn coating, whereas the thick and broken vertical arrows show the direction of significant and negligible atomic transport, respectively. The vertical channel (onto which the stresses are shown acting) in (b) represent a columnar grain.

FEA of the bimetallic strip with geometry similar to that of the actual sample was also employed using ANSYS, a commercial FEA software package, for quantification of the stress state as well as the stress gradient across the coating thickness imposed due to the 3-point bending. The FEA was performed by assigning time independent elastic-plastic property for Sn and assuming brass to be elastic.

After electro-deposition and loading the sample into 3-point bending fixture, it was stored in a hot air oven at 50 °C and whisker density as well as the $Cu_6Sn_5$ IMC layer was observed after 7 days. IMC layer was observed using both the cross-section imaging and chemically etching away the Sn coatings from the substrate. The sample for cross-section examination was prepared using dual beam focused ion beam (FIB) machining, whereas Sn coating was etched away using 7 parts o-nitro-phenol and 1-part NaOH warm solution. Whisker density in the samples was evaluated at several distances away from the center (i.e., maximum deflection point) and correlated with the magnitude of stress and stress gradient that exist at the particular location.

**3.    Results**
3.1    <u>Substrate curvature measurements</u>

**Figure 2** shows the evolution of bulk stress in a representative Sn coating aged at room temperature, as estimated using **Eq. 1**. Careful observation of **Fig. 2** readily reveals the following: (i) the residual stress in the Sn coating immediately after deposition was compressive, (ii) the initial compressive stress in the Sn coatings relaxed rapidly during initial a few hours of aging after the deposition, (iii) thereafter, stress became progressively more compressive with aging, and (iv) the compressive stress showed saturation near -12 MPa (which is close to yield strength of Sn) after 400 h of aging.  Interestingly, whiskers were observed in this sample just after 7 days (i.e., ~180 h), suggesting that whiskering did not completely stop the stress built-up in Sn coatings. However, saturation in stress or steady state stress after 400 h suggests that once very large number of whiskers were formed, the stress generation due to IMC formation at the substrate-coating interface and the stress relaxation due to whisker formation balanced each other. It should be noted that creep of Sn may also contribute to the stress relaxation in Sn coating. However, a saturation in the stress confirms that IMC growth induced generation of compressive stress in Sn coating remained significant throughout the storage of the coating, otherwise the coating should approach a stress-free state.  It should be noted that the measured stress corresponds to the entire Sn coating (i.e., both bulk and surface). However, it is expected that the average value will be more affected by the volume of the material away from the top surface and hence the stress shown in **Fig. 2** can be assumed to represent the bulk stress.

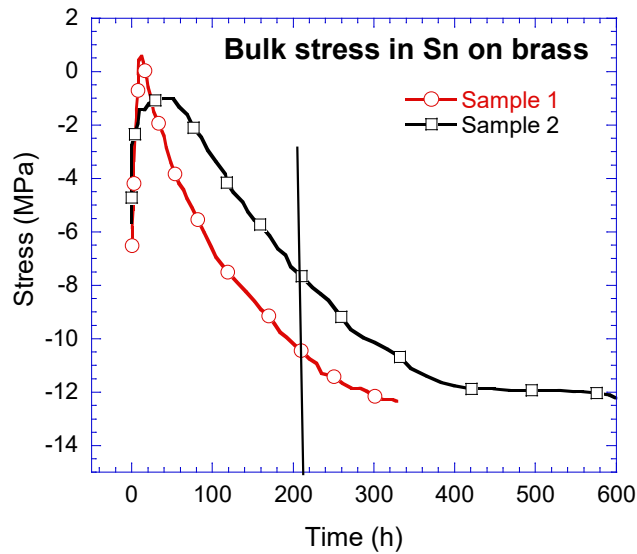

Fig. 2: Bulk stress evolution in two 4 μm thick Sn coating electro-deposited on 200 μm thick brass substrate with time, as measured using the laser-optics based curvature measurement setup. The vertical line shows the approximate time at which the first set of whiskers started to grow in these coatings. Samples were stored in the ambient condition.

It is important to note that when Sn is deposited on brass substrate, a scaloppy layer of $Cu_6Sn_5$ IMC grows with time near the coating-substrate interface. Therefore, the measured curvature is proportional to $\sigma_{Sn}.t_{Sn}+\sigma_{IMC}.t_{IMC}$, where $\sigma_{Sn}$ and $\sigma_{IMC}$ are the stresses in Sn coating and IMC layer, respectively, and $t_{Sn}$ and $t_{IMC}$ are the thicknesses of Sn coating and IMC layer, respectively. However, it important to notice that the interfacial IMC layer is not uniform and continuous. In fact, IMC scallops only grow along grain boundaries and triple points, and coarsen with aging to form a continuous network IMC along the grain boundaries after very long time. The effect of such a discontinuous layer on curvature and hence stress in film was first modeled by Beuth [34] by assuming multiple cracks extending through the thickness of the film. Later, Hutchinson developed a modified Stoney's equation by incorporating a Hutchinson factor that depends on the ratio of spacing between adjacent cracks, $d$, (which will be distance between IMC scallops) and layer thickness, $t_f$, (which will be equal to the height of IMC scallops in this case) [35]. When $d \gg t_f$ the Hutchinson factor, $H$ tends to 0 and normal Stoney's equation is recovered; however, when $d \ll t_f$, then $H$ becomes equal to 1 and the induced curvature is small and can be neglected, leading to a stress-free film [35]. The average distance between IMC scallops (i.e., $d$) and height of IMC scallops (i.e., $t_f$) was measured using SEM image analysis of substrate after selectively etching the Sn coating, and it was observed that $H$ was close to 1. Therefore, according to the aforementioned analysis, the curvature due to IMC layer could be assumed to be negligible, and hence the stress evolution shown in **Fig. 2** can be directly attributed solely to the evolution of stress in the Sn coating.

### 3.2 Stress measurement using X-ray diffraction

**Figure 3a** shows the peak profile of (501)-planes obtained at different angles of incidence. The vertical arrow in **Fig. 3a** shows the expected position of the peak of (501)-reflection in unstrained crystal, and any deviation from this arrow can be attributed to the strain in the sample. **Fig. 3** clearly shows a significant shift in the peak position of (501)-reflection in Sn coating after 5 days of aging. Subsequently, the strain values at different incidence angles were plotted against $sin^2\psi$ to calculate the average stress in the top 0.76 μm of the coating. As directed by **Eq. 3, the** strain values at different incidence angles, $\omega$, and therefore at different $\psi$ tilts, were plotted as function of $sin^2\psi$ (see **Figure. 3b**). As shown in **Fig. 3b**, a reasonably linear fit was obtained between the strain and $sin^2\psi$ values. The stress was calculated from the slope of the linear fit using **Eq. 3**. It should be categorically noted that the value of stress reported following this method is the average stress near the surface of Sn coatings (i.e., top 0.7 μm of the coating). Subsequently, stress near the surface region was measured using the aforementioned technique after different aging time up to 30 days after electro-deposition.

(a) (b)

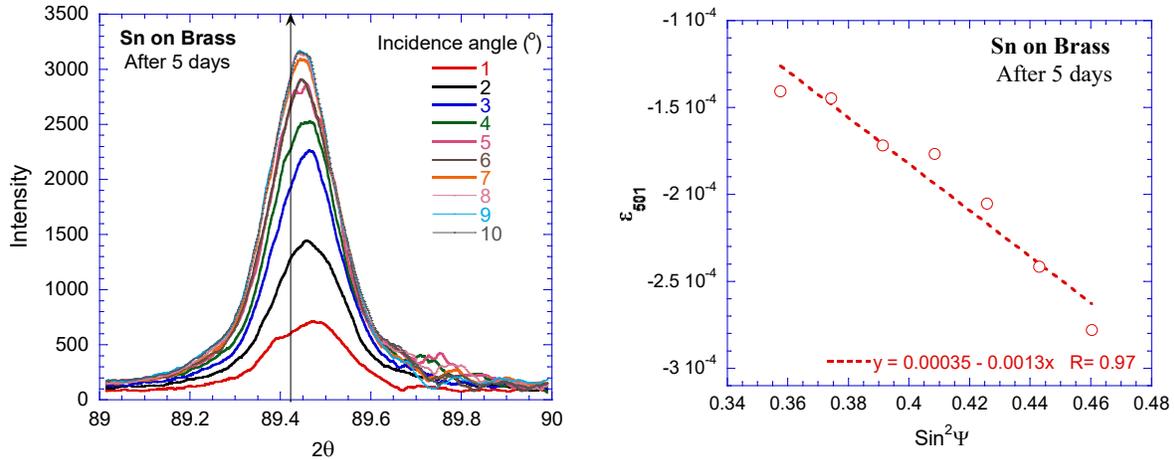

Fig. 3: (a) Peak profile of (501)-reflections of Sn coating obtained at different incidence angles, ω, shown in the legend. The data was collected 5 days after the electro-deposition. (b) Strain versus $sin^2\psi$ plot for the Sn coating deposited on brass as recorded after 5 days since the electro-deposition. The legend shows equation obtained from the best linear curve fit, along with the curve fitting parameter, $R$.

**Figure 4** shows the $Sin^2\psi$ plot of the same coating, whose data is shown in **Fig. 3,** as recorded at different time along with the variation of stress with time. It can be readily observed from **Fig. 4** that the stress in the as-deposited samples was compressive, which became less compressive (i.e., more tensile) with aging time. This suggests that whisker growth is a surface relief process, such that surface stress became less compressive continuously during whisker growth. A comparison of **Figs. 2** and **4b** clearly suggests that the stress gradient in the Sn coating between the bulk and the top layer of the coating became more negative (i.e., negatively increased) with the storage time. The implication of this observation on whisker growth will be discussed later.

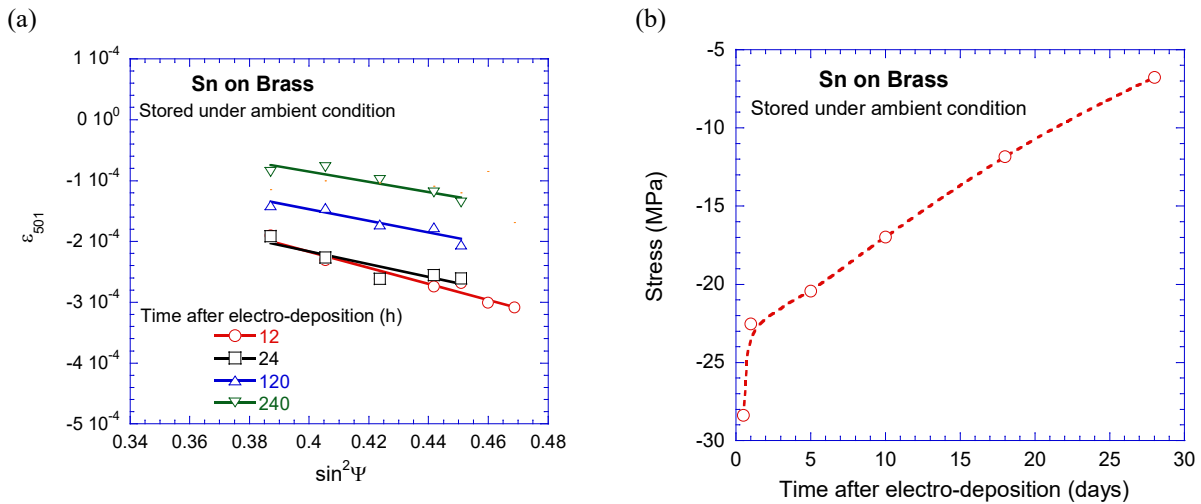

Fig. 4: (a) A few $Sin^2\psi$ plots of the Sn coating on brass substrate as obtained at various time periods after the electro-deposition, and (b) evolution of average surface stress with aging time after electro-deposition.

It should be noted that the effect of crystallographic texture and grain interaction between polycrystalline grains imposing constraints on the neighboring grain is not considered in this XRD stress analysis. The analysis assumes rotationally symmetric equal biaxial stress in Sn and macroscopically isotropic specimen. Therefore, the XRD stress values are expected to be higher than the actual values. Furthermore, as whisker starts to a grow at a particular location, the stress state in the vicinity of the whisker grain will change. Such local variation in the stress field was not captured using the performed XRD (as well as curvature measurement) analysis. Nevertheless, it captures the evolution of average (or global surface) stress in the Sn coating with the aging time.

3.3 <u>Effect of external stress on whisker growth</u>

As discussed in **Section 2.3**, an external stress along with stress gradient was imposed on the Sn coated brass strips using a 3-point bending setup. The stress variation in the Sn coated brass strip at various distances away from center (i.e., point of maximum deflection) is shown in **Figure 5**. A saturation in stress in the Sn coating if Sn is allowed to deform plastically with negligible hardening can be observed. Henceforth, the stress values, at different locations along the length of a Sn coated strip reported in this study, are the "predicted" values using FEM model incorporating plasticity.

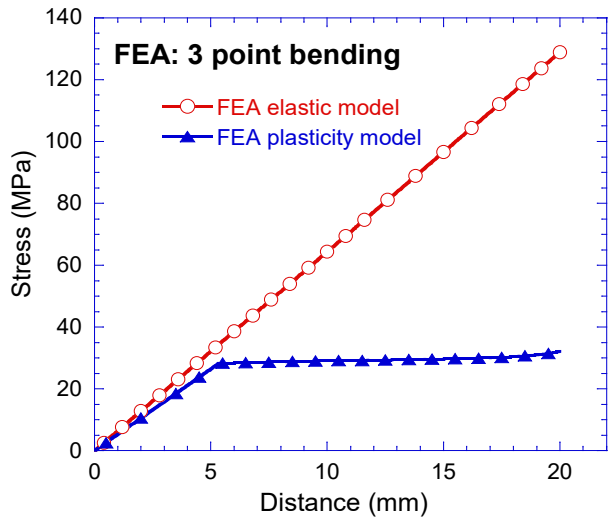

Fig. 5: Variation of magnitude of the stress in the Sn coating, subjected to 3-point bending, as a function of distance from the center (or point of maximum deflection), as predicted by FEA. The stress values in Sn vary linearly and becomes very high if Sn is not assumed to be an elastic-plastic material.

**Figure 6** shows a few representative micrographs of the Sn coatings stored at 50 °C for 7 days under applied tensile and compressive stresses of similar magnitude. It is evident from **Fig. 6** that the whisker density was much higher in the Sn coating under tensile stress. Also, not only the number density of whiskers was larger in the Sn coatings under tensile stress, but also the average length of the whiskers was significantly higher as compared to the Sn coatings loaded under compressive stress of equal magnitude.

(a)     (b)

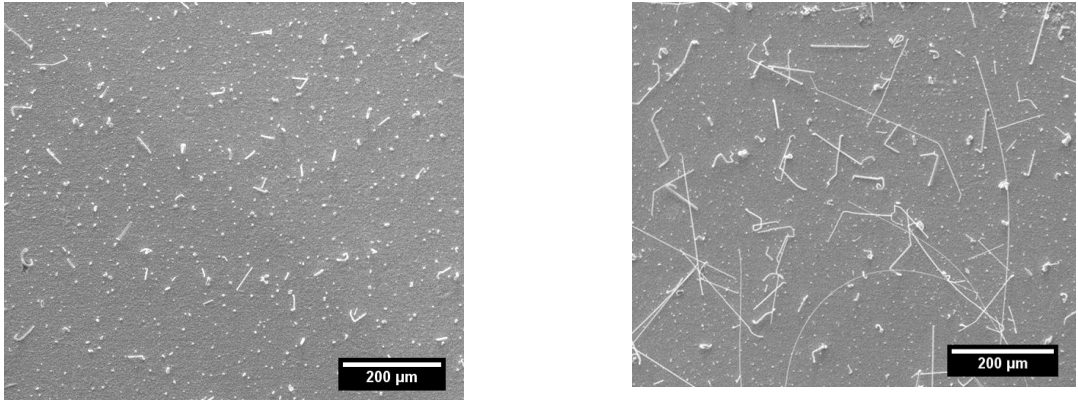

Fig. 6: Representative micrographs, obtained using scanning electron microscope, of Sn coatings subjected to external (a) compressive and (b) tensile stresses of similar magnitudes. The stresses were applied using a 3-point bending setup.

**Figure 7** shows whisker density, measured as number of whiskers per unit area, as a function of magnitudes of the stress and the stress gradient. Whisker density was calculated at different fixed nominal distances (e.g., 0, 5, 10, 15, 20 and 22.5 mm) from the center. A region of 0.8 mm length such that the region lies ±0.4 mm on each side of the nominal distance mentioned above was observed for whisker density calculation. Furthermore, the whisker density values reported in **Fig. 7** is the average whisker density over the region mentioned above and the stress as well as stress gradient shown in **Fig. 7** corresponds to the FEA predicted value at the center of the 0.8 mm wide region wherein the whisker density was measured. It is evident from **Fig. 7a** that, consistent with **Fig. 6**, the whisker growth was very rapid when Sn coating was under tensile stress. Also, the whisker density increased monotonically with the tensile stress and slightly decreased with increase in the magnitude of the compressive stress. Interestingly, the effect of increase in the tensile stress on whisker density was significantly more prominent as compared to that of the compressive stress. Similarly, **Fig. 7b** shows that whisker density monotonically increased with the magnitude of the negative stress gradient across the thickness of Sn coating imposed due to bending.

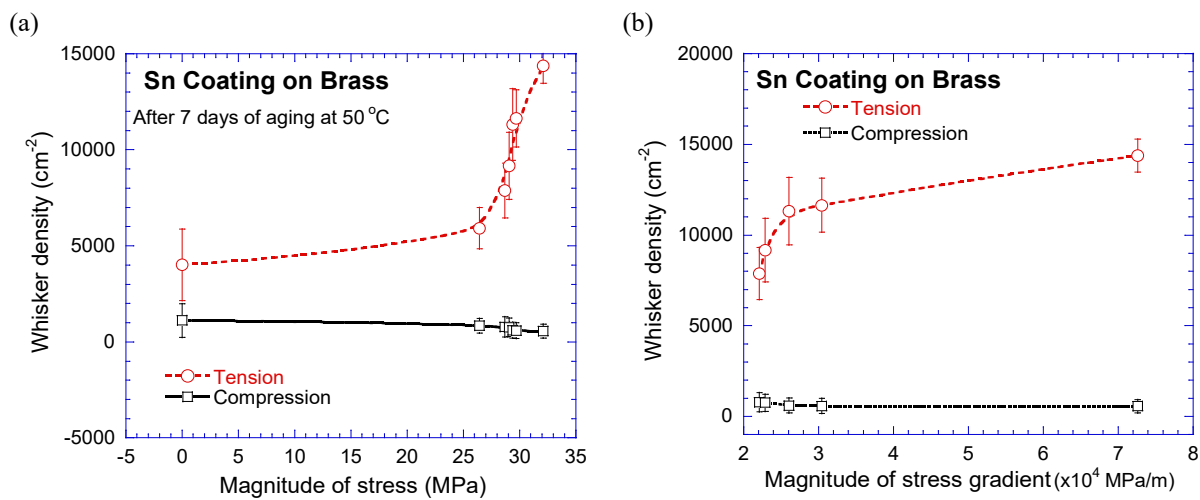

Fig. 7: Whisker density observed in the samples loaded using a 3-point bending setup at 50 °C after 7 days of electro-deposition as function of the magnitude of the (a) stress and (b) stress gradient across thickness of Sn coating.

It should be noted that the externally applied stress (and hence stress gradient across the coating thickness due to bending) might also affect the diffusion of Cu and Zn atoms from the substrate into the Sn coating [39] and hence the growth kinetics of interfacial IMC layer in the coatings under compressive and tensile regions is expected to be significantly different. Therefore, the IMC growth in the samples loaded using 3-point bending setup was monitored. **Figure 8** shows micrographs revealing the cross-section of the Sn coatings under compressive and tensile stresses of similar magnitude. Careful observation of **Fig. 8** reveals that the thickness of the interfacial IMC layer in the Sn coatings under the tensile stress was $1.6 \pm 0.4$ μm, while it was $1.5 \pm 0.4$ μm in the coatings under the compressive stress of similar magnitude. Therefore, the IMC thickness, as calculated using analysis of the cross-sectional micrographs, statistically did not show noticeable dependence on the sign of stress or stress gradient.

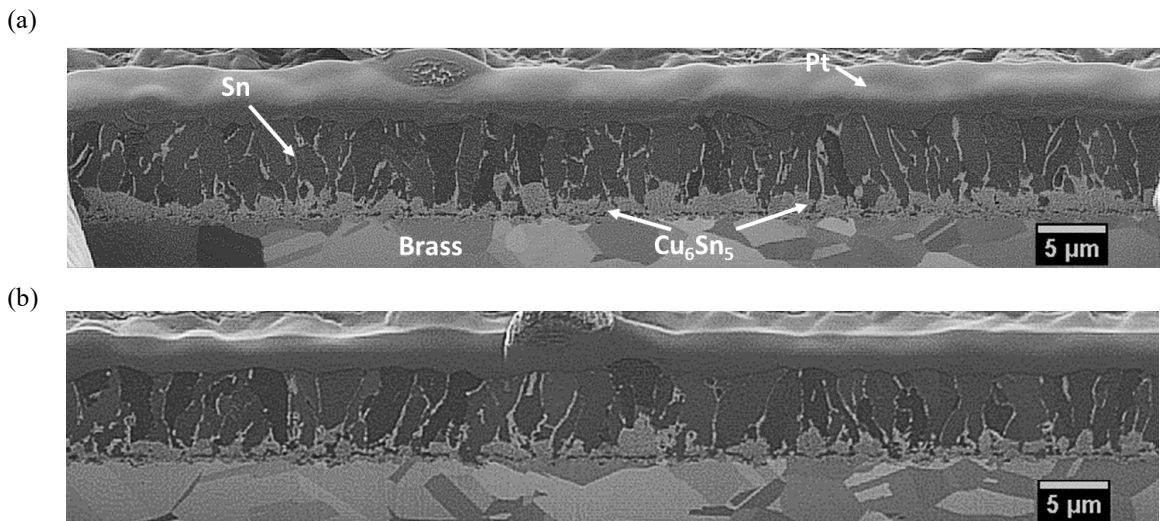

Fig. 8: Representative micrographs showing the cross-section of Sn coatings subjected to (a) tensile and (b) compressive stresses of similar magnitude.

**Figure 9** shows the top view of the interfacial IMC layer as obtained after the selective etching of the Sn coating. **Fig. 9** clearly shows that the IMC formed on the tensile side of the coating (substrate) was significantly denser as compared to the IMC layer formed in Sn coating on the compression side. The area fraction of IMC, calculated using image analysis, on the tensile side and the compression side were 0.65 and 0.43, respectively. Therefore, it can be inferred that although the sign of the stress in sample loaded under 3-point bending did not affect the overall thickness of the IMC layer, the overall volume of IMC was drastically affected by the nature of stress, wherein larger tensile stress led to an increase in the volume of the interfacial IMC layer. **Fig. 9** also reveals that the IMC layer grown under externally applied stress retained its scallopy nature.

(a)                                                        (b)

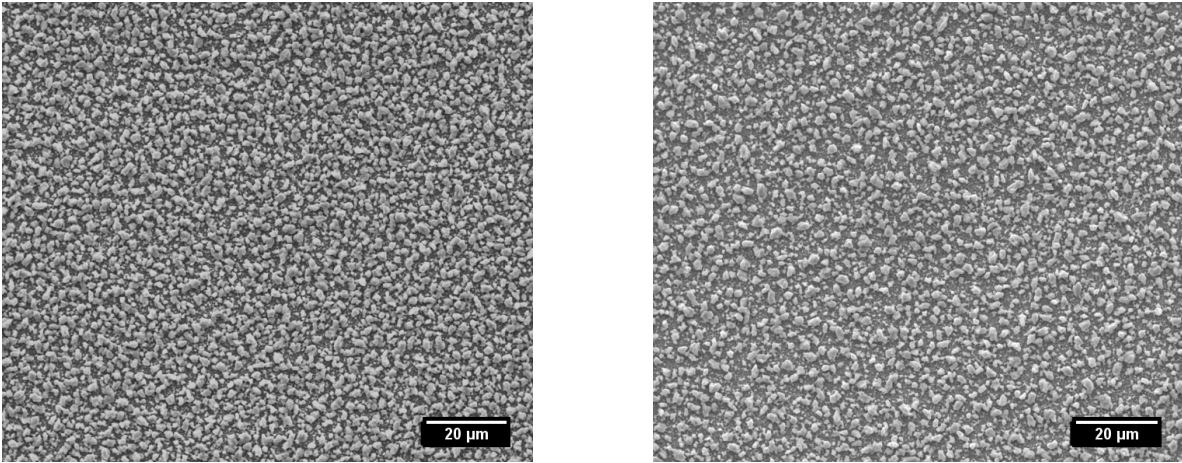
Fig. 9: Top view of IMC layer after selectively removing the Sn coatings exposed under (a) tensile and (b) compressive stresses of similar magnitudes.

## 4. Discussion
### 4.1 Effects of stress and stress gradient on whisker growth

The residual bulk stress in as-deposited Sn coatings was compressive (see **Fig. 2**). Assuming the stress in Sn coating due to growth of IMC layer to be insignificant immediately after electro-deposition, the initial residual stress in the coating may originate from mismatch in the coefficients of thermal expansion (CTE) and/or lattice misfit between the coating and the substrate or it may be inherent to the electro-deposition process. Since the Sn coating was electro-deposited at constant temperature of 40 °C and then aged under ambient condition (i.e., at a nominal temperature of 25 °C) in the experiments wherein evolution of stress was recorded, the coating was expected to be under compressive stress immediately after deposition due to CTE mismatch between Sn and brass. It should be noted that Sn has higher CTE compared to the brass substrate and hence the residual compressive stress built up in Sn can be estimated using the following simplified (1-D) equation:

$$\sigma_{Sn} = M.\Delta\alpha.\Delta T \tag{5}$$

where $M$ is equal to $E/(1-\nu)$ and $\Delta\alpha$ is the difference in coefficient of thermal expansion between Sn and brass. Now, taking the elastic modulus of 50 GPa for Sn [40], the residual compressive stress in Sn due to the 15 °C temperature difference between the electro-deposition bath temperature and the room temperature will be equal to ~5.2 MPa, which is very close to the bulk stress value in Sn coating measured using laser-optics based setup immediately after electro-deposition (see **Fig. 2**).

It was also observed that compressive stress in Sn coatings relaxed rapidly in initial a few hours of room temperature aging and thereafter stress became increasingly compressive and ultimately saturated around -12 MPa (see **Fig. 2**). The initial compressive stresses, when significant volume of IMC was not formed, quickly relaxed by creep and other recovery process. This is expected because the room temperature is a very high temperature for Sn (~0.58 $T_m$, where $T_m$ is the melting temperature). With continued room temperature aging the columnar grain boundaries of Sn would get pinned both at the top and the bottom due to oxide layer and IMC layer formed at the grain boundaries terminating at the free surface and the substrate, respectively. The formation of oxide on the free surface of Sn does not allow stresses to be efficiently relaxed by creep processes. Therefore, due to the continuous growth of $Cu_6Sn_5$ IMC along the grain boundaries of Sn terminating on the brass substrate, the stress in Sn coating becomes progressively more compressive, as shown in **Fig. 2**. The specific volume of the $Cu_6Sn_5$ is larger (~0.22%) than the supersaturated solid solution Sn and Cu [14] and hence formation of $Cu_6Sn_5$ at the Sn grain boundaries would generate compressive stress in the adjacent Sn grains [18]. Interestingly, the built-up of compressive stress in Sn eventually showed saturation after prolonged aging (>400 h). This saturation in stress can be achieved by the formation of enough number of whiskers from the Sn coating such that incremental increase in the compressive stress due to IMC formation could be quickly relaxed by the incremental growth of whiskers on the surface of Sn coating. It should be noted that the growth rate of IMC slows down with time, $t$ (as $h_{IMC} \sim t^n$, where $n < 1$, e.g., ½, etc.). This reflects also in the slowing down of the growth rate of whiskers in Sn coatings with time [43]. This balance between stress generation due to IMC and stress relaxation due to whisker growth seems to be manifested as a steady state compressive stress in Sn after prolonged aging.

Quite interestingly, as shown in **Fig. 4b,** the stress measurement using XRD showed the opposite trend in the stress evolution, wherein the initial stress in the Sn coating was compressive and the stress became less compressive (or, more tensile) with aging. It should be noted that stress measured using XRD was the average stress in only the top 0.7 μm layer of the Sn coating. Now, it is known that under ambient condition, Sn forms a stable oxide $SnO_2$. Since $SnO_2$ has higher molar volume than the Sn [41], formation of $SnO_2$ along Sn grain boundaries terminating on the free surface, where diffusion of oxygen is faster [41], will induce compressive stress near the surface of the Sn coating. Since the oxide layer is very thin, the compressive stresses will only be induced in the top most regions. Now, the decrease in the compressive stress in the top layer with time can be explained as follows: (i) breaking of $SnO_2$ layer along grain boundaries of Sn due to nucleation and growth of whiskers, and (ii) replacement of $SnO_2$ by more stable ZnO. ZnO forms due to the fast diffusion of Zn atoms from the brass substrate along the grain boundaries of Sn and the subsequent reaction with $SnO_2$. The fast diffusion of Zn was confirmed using electron probe microanalysis (EPMA) in one of our previous works [43] and also in a few other studies [44, 45], where considerable concentration of Zn was noticed near the free surface of the Sn coating. The molar volume of ZnO is 14.5 $cm^3$/mol [42], which is considerably smaller than the molar volume of $SnO_2$, which is 21.7 $cm^3$/mol [41]. Therefore, replacing $SnO_2$ with ZnO will relax the existing compressive stress near the surface of the coatings. In summary, formation of ZnO by reaction between $SnO_2$ and Zn and breaking of the oxide layer explains the decrease in the compressive stress near the surface of the Sn coatings with time. Therefore, it can be inferred that mechanisms of the stress generation and the stress relaxation at the coating-substrate interface and at the free surface in the Sn coating, respectively, are fundamentally different.

**Figure 10** compares the evolution of the bulk stress and the surface stress in Sn coatings, as individually shown in **Fig. 2** and **4b**, respectively. It can be readily observed from the **Fig. 10** that the surface stress started to become less compressive relative to the bulk with time and, on an average, it became more tensile than the bulk stress after 420 h (~17 days), as marked by solid vertical line. Therefore, while it is expected that Sn atoms would tend to migrate from the surface of the coating towards the coating-substrate interface in the beginning of the aging, the transport of Sn atoms would eventually become upward towards the surface, thereby supplying the needed mass flux at the root of whisker needed for its growth. It should be noted that the broken vertical line drawn in the **Fig. 10** shows the instance when the whisker growth was first observed (i.e., after ~180 h), thereby confirming existence of an incubation period during which stresses are built-up in Sn coating. Although the surface is overall (or an average) more compressive than the bulk at the time of growth of first set of whiskers, the broken oxide layer along the grain boundaries of Sn near surface can provide a stress-free location for diffusion of Sn atoms. A strong directional negative stress gradient towards such stress-free spot would be locally established as soon as a small patch of the oxide breaks and towards which Sn atoms can be transported for whiskers to grow. Since the employed stress measurement techniques are not sensitive to the local stress field and reflect only the average stress, growth of whiskers due to local variation of stress field could not be captured in this study. However, it can be expected that as soon as the bulk on an average became more compressive than the surface, significant mass flux of Sn atoms would occur from the bulk to the surface, thereby accelerating the growth of whiskers and proffering spontaneous growth of whiskers over long period of time. Following an accelerated period of whisker growth, the whisker growth rate will slow down once the built up of compressive stress in bulk of Sn gets saturated. Therefore, by observing stress profiles in the bulk and in the region in vicinity of the surface of the coating, as shown in **Fig. 10**, the existence of an incubation period for whisker growth, followed by a period of accelerated rate of whisker growth followed by a slowdown in whisker growth can be explained.

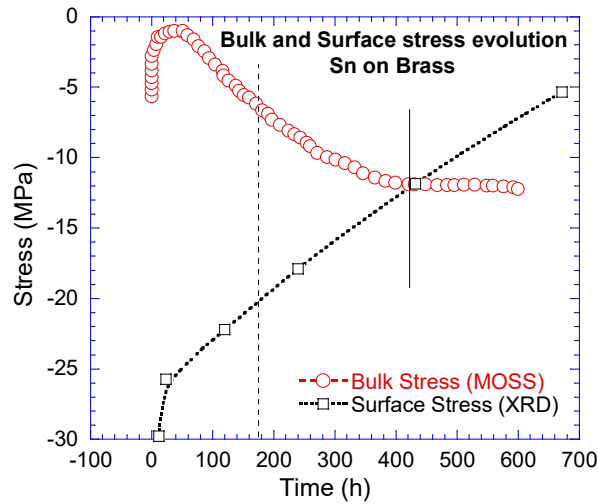

Fig. 10: The evolution of bulk stress measured by substrate curvature technique and the surface stress measured using glancing angle XRD in Sn coating on brass substrate with time. The broken and solid vertical lines denote the onset of first set of whiskers and instance when the bulk became on an average more compressive than the surface.

In addition, the evolution of bulk stress and surface stress systematically captured here can explain few widely reported observations related to the whisker growth in Sn coatings: (i) the evolution of stress and the stress gradient between the surface and the bulk can explain different incubation periods observed in the previous works [15,49] for different Sn coatings, (ii) the existence of a positive stress gradient up to prolonged aging time favoring transport of atoms from surface towards the interface can explain depletion zones or subsiding grains reported in literature [30,50], and (iii) cross-over of bulk stress and surface stress producing negative stress gradient favoring mass transport of atoms from compressive region near IMC towards whisker root can explain continued growth of whiskers up to prolonged time.

## 4.2 Effect of externally imposed stress and stress gradient on whisker growth

As discussed in the previous section, both the surface and the bulk stresses and the produced stress gradients play critical roles in whisker growth. The externally applied stress, as shown in **Fig. 1,** will also affect the stress state in the top layer of the coating, so that the Sn coating subjected to the outward bending (i.e., convex side) will have the maximum tensile stress at the surface and the inward bent (i.e., concave side) coating will have the maximum compressive stress at the coating surface. Therefore, a negative stress gradient, necessary for whisker growth, was inherently imposed in the Sn coatings that are outward bent, i.e., which are under tensile stress. The direction of such a stress gradient will be reversed for the inward bent coatings. This makes the outward bent coatings, i.e., the coating under tensile stress, more favorable for whiskers to grow.

In addition, it was shown in **Figs. 8** and **9** that although the thickness (i.e., growth rate) of interfacial IMC layer was similar in the coatings loaded under tensile and compressive stresses, the volume of IMC formed (i.e., nucleation rate of IMC) in the coatings under tension was significantly more compared to the coatings under compression. The enhanced nucleation of IMC in the Sn coatings loaded under tensile stress can be attributed to the fact that most of the columnar grain boundaries of Sn terminating on the substrate experienced a tensile opening force. This enhances the diffusion of Cu atoms and subsequent nucleation of IMC as compared to the Sn coating loaded under compressive stress, wherein most of the boundaries experience a compressive stress (see **Fig. 1b** for schematic illustration). As soon as the interfacial IMC layer starts to grow, the negative stress gradient in the outward bent coating will be further augmented by to the additional influence of the intrinsically generated compressive stress due to the formation of interfacial IMC layer. Interestingly, the externally applied tensile stress in the outward bent coating may also break the (brittle) surface oxide layer at several spots, thus increasing the probable locations for whisker growth as well as outward stress gradient. These corroborate with the observation that the length of whiskers as well as the growth rate of whiskers in outward bent coatings, which were under tensile stress, was significantly higher as compared to the inward bent coatings.

It should be noted that similar observation pertaining to the enhanced whisker growth under tensile stress was also reported by Chen et al. [15], Crandall et al. [29], and Cheng et al. [30]. However, explanation for the observations was lacking. This is the first report that presents a systematic variation in the whisker density as function of tensile as well as compressive stress, and, more importantly, also explains the whisker growth even under tensile stress by taking stress gradient into account. It is imperative from the evidences shown in this study that the negative stress gradient, rather than the of the stress, seems to dominantly govern whisker growth.

## 5. Conclusions
- Stress state in the Sn coatings on brass was studied using two techniques, viz. optical substrate curvature measurement using laser-optics and glancing X-ray diffraction. The curvature technique allowed measurement of average bulk stresses in Sn and its evolution with aging time. Similarly, the glancing angle X-ray diffraction allowed measurement of stress in the top ~0.7 μm layer near the free surface of the coating and its evolution with aging time.

- The surface and the bulk stresses in Sn coating evolved differently with time. The bulk stress in the Sn was compressive immediately after electro-deposition; however, it rapidly relaxed and thereafter progressively became more compressive due to IMC growth until stress build-up due to IMC formation was balanced by stress relaxation by whisker formation. This led to saturation in stress values. On contrary, stress in the Sn layer near the top surface of the coating became less compressive (i.e., more tensile) with time.
- Initially, a positive stress gradient (i.e., coating substrate interface was more tensile than the surface) was observed in the Sn coating which reversed its sign after a certain aging period (i.e., 400 h at room temperature), enabling dramatic increase in the transport of Sn atoms from the highly-compressed Sn layer near IMC to the whisker root.
- Whisker density increased monotonically with externally imposed tensile stress and negative stress gradient. The enhanced whisker growth was due to the combined effect of the ease of breaking of the surface oxide layer at many spots and a negative stress gradient towards such spots inherently imposed in such outward bent (or tensile loaded) Sn coatings.
- The externally imposed tensile stress along with the negative stress gradient accelerated the IMC growth such that density of IMC observed in Sn coatings under tensile stress with native stress gradient was higher than the coatings under compressive stress with positive stress gradient.

**Acknowledgements**

This work was financially supported by Indian Space Research Organization (ISRO) through Space Technology Cell at Indian Institute of Science, Bangalore (Grant # ISTC 0367).